%% file: arxiv.tex
\documentclass[journal,twoside]{IEEEtran}

\usepackage{cite}
\usepackage{amsmath,amssymb,amsfonts}
\usepackage{algorithmic}
\usepackage{graphicx}
\usepackage{textcomp}
\usepackage{xcolor}
\usepackage{hyperref}
\usepackage{braket}
\usepackage{subcaption}
\usepackage[version=3]{mhchem} 

\title{Harnessing a 256-qubit Neutral Atom Simulator for Graph Classification
}

\author{
\IEEEauthorblockN{
Edoardo Giusto\IEEEauthorrefmark{1},
Gabriele Iurlaro\IEEEauthorrefmark{2},
Bartolomeo Montrucchio\IEEEauthorrefmark{2},\\
Alberto Scionti\IEEEauthorrefmark{3},
Olivier Terzo\IEEEauthorrefmark{3},
Chiara Vercellino\IEEEauthorrefmark{2}\IEEEauthorrefmark{3},
Giacomo Vitali\IEEEauthorrefmark{2}\IEEEauthorrefmark{3}\IEEEauthorrefmark{4},
Paolo Viviani\IEEEauthorrefmark{3},
}\\
\IEEEauthorrefmark{1}\textit{University of Naples, Federico II}, Napoli, Italy\\
\IEEEauthorrefmark{2}\textit{\textit{DAUIN}, Politecnico di Torino}, Torino, Italy\\
\IEEEauthorrefmark{3}\textit{Fondazione LINKS}, Torino, Italy\\
\IEEEauthorrefmark{4}\textit{giacomo.vitali@linksfoundation.com}
}

\begin{document}
\maketitle

\begin{abstract}
Neutral atom platforms are analogue quantum simulators that offer the possibility to map graphs onto a 2D qubit register using programmable Rubidium atoms arrays, whose valence electrons' energy state is used as qubits, using optical tweezers. This makes it possible to implement algorithms for solving graph combinatorial optimization and Quantum Machine Learning (QML) tasks, such as graph classification. However, the restrictions of real hardware, as well as the very low number of publicly available machines, make such implementation non-trivial. In this work, we manage to compute the Quantum Evolution Kernel (QEK) to extract the features from graphs of the  \textit{PROTEINS} dataset using the 256-qubits Aquila platform (available through AWS) and then we apply classical Machine Learning (ML) techniques for the final classification. The method is benchmarked against classical kernels, resulting in slightly better performance, proving the effectiveness of the method, even in the case of a noisy quantum simulator.
\end{abstract}


\input{src/01_intro.tex}
\input{src/02_related_work.tex}
\input{src/03_methodology.tex}
\input{src/04_results.tex}
\input{src/05_conclusion.tex}

\section*{Acknowledgment}
We would like to thank AWS for the research credit grant and QuEra for the support and useful discussions.
% \section*{References}

\bibliographystyle{unsrt}
\bibliography{refs}

\end{document}

%% file: src/01_intro.tex
\section{Introduction}\label{sec:intro}

Quantum Machine Learning is one of the most recent and interesting fields of application of Quantum Computing (QC). The wide range of QC technologies in rapid development in recent years, all with their own set of characteristic, such as relaxation times, qubit connectivity, gate fidelities, etc., contribute to create a complex scenario where it is not straightforward to choose the best suited combination of algorithm and hardware platform for a specific Machine Learning (ML) task.

In this sense, quantum simulators are a QC typology very different from most well-known platforms, such as IBM superconducting machines, as they operate in analogue mode, i.e., the qubit state evolves following the machine Hamiltonian which depends on the specific technology employed. This typology of quantum computers allows to natively explore quantum system, such as spin models, which have applications especially in the field of quantum chemistry and new materials development.

Among simulators, neutral atom platforms \cite{Henriet_2020} are rapidly developing thanks to some peculiarities.
These machines operate optical tweezers to arbitrarily position alkaline atoms \cite{Barredo_2018, Nogrette_2014}, commonly Rubidium (\ce{^{87}Rb}), in a 2D plane, while the valence electron's energy state serves as the two-state qubit, for example, the ground state $\ket{g}$ as $\ket{0}$ and a high-energy state (e.g., $70S$ for \ce{^{87}Rb}), called Rydberg state $\ket{r}$, as $\ket{1}$.
Furthermore, the connectivity of the qubits can be tuned due to the Rydberg blockade effect, which creates entanglement among the qubits within the Rydberg radius $r_b$. In fact, the valence electron can be excited to a high-energy level through a dedicated Rydberg laser. In formal terms, the Hamiltonian of a typical neutral atom simulator is: 

\begin{align}\label{hamiltonian_eq}
    \hat{\mathcal{H}}(t) = & \sum_{i=1}^{n} \frac{\hslash \Omega_i(t)}{2} ( e^{i\phi(t)} \ket{0_i}\bra{1_i} + e^{-i\phi(t)} \ket{1_i}\bra{0_i} ) \nonumber\\ 
    & - \sum_{i=1}^{n} \hslash \Delta_i(t) \hat{n}_i + \sum_{j>i} \frac{C_6}{ \lvert \overrightarrow{x}_i - \overrightarrow{x}_j \rvert ^{6}} \hat{n}_i \hat{n}_j ,
\end{align}

where $\Omega_i$ and $\Delta_i$ are the time-dependent Rabi frequency and detuning, respectively, acting on the i-th atom (both measured in $rad/\mu s$), $\phi$ is the phase of the Rabi drive, $\overrightarrow{x}_i$ is the position of the i-th atom, $\hat{n}_i$ is the operator $\ket{1_i}\bra{1_i}$ which counts the atoms in the excited state and $C_6$ is the Rydberg interaction coefficient, whose value depends on the chosen Rydberg level. The last term, the Van der Waals interaction, is often rewritten using $V_{ij} = \frac{C_6}{ \lvert \overrightarrow{x}_i - \overrightarrow{x}_j \rvert ^{6}}$.
With this formalization, the Rydberg radius of a neutral atom can be defined as $r_b = (\frac{C_6}{\hslash\sqrt{\Omega^2 + \Delta^2}})^{1/6}$.
When two atoms atoms are inside $r_b$, we have that $\Omega \ll V_{ij}$ $\Delta \ll V_{ij}$, and they are in the Blockade regime \cite{Ga_tan_2009, Jaksch_2000}, i.e., the $\ket{11}$ is energetically prohibited and therefore suppressed.
The programmability of the qubit array \cite{Morgado_2021} and the connectivity characteristic of these devices allow the mapping of graphs to the register by associating graph nodes to qubits, which share an edge when $ \lvert \overrightarrow{x}_i - \overrightarrow{x}_j \rvert < r_b $. It is important to note that, while each atom can be theoretically fully addressable, in practice the available neutral atom platforms are at the moment restricted to global pulses, i.e., the $\Omega$ and $\Delta$ terms can be taken outside of the summations of Eq. \ref{hamiltonian_eq}. This means that qubits in the register represent Unit-Disk (UD) graphs. For this reason an embedding step is necessary, as detailed later.\\
The features just mentioned allow one to apply quantum algorithms to several problems from different domains, such as graph combinatorial optimization and ML techniques.
In this paper, we describe the development and implementation of a QML technique for graph classification, where we use the time evolution determined by the neutral atom Hamiltonian to extract features from a test dataset's graphs, compute the Quantum Evolution Kernel (QEK) defined in \cite{henry2021quantum}, and use a Support Vector Machine (SVM) to classify them. Finally, the method is benchmarked against a well-known classical kernel.\\
In particular, we perform a binary classification of proteins in the PROTEINS dataset~\cite{dobson2003distinguishing,borgwardt2005protein} in the two-class labeled as \textit{enzymes} and \textit{non-enzymes}.
To this end, we used the Aquila 256-qubit neutral atom simulator built by QuEra Computing \cite{wurtz2023aquila}, available through Amazon Web Services (AWS) Braket.\\
This paper is structured as follows: first we report the context of this work, together with references to relevant bibliography. In Section \ref{sec:methodology} we described the methodology of the developed classification procedure and the characteristics of the PROTEINS dataset, as well as the preprocessing steps needed for the QEK implementation on neutral atom simulators. In Section \ref{subsec:emu} we report the results obtained by emulating with classical resource the quantum evolution on a small subset of the considered dataset and detail the pulse optimization procedure performed. Finally, in Section \ref{sec:aquila} we compare the classification results obtained using Aquila with a well-known classical kernel and summarize the results of the work in Section \ref{sec:conclusion}.

%% file: src/02_related_work.tex
\section{Related Work}

Graphs-based data are widely used in scientific and industrial applications. From telecommunication \cite{POURHABIBI2020113303} to biology \cite{genes, proteins_dl} and social science \cite{Correa2011}, they allow describing interactions among data.\\
However, given their intrinsic complexity, algorithms that consume graph data are usually very resource-hungry. The exponentially large Hilbert space offered by quantum computers has prompted the interest of scientists and domain experts for this type of application, especially in the field of QML \cite{Schuld_2019, PhysRevA.101.032314, Kishi_2022}. Some examples are quantum convolutional neural networks\cite{Cong_2019, zheng2021quantum, verdon2019quantum}. In the field of quantum kernels, a version developed for universal gate quantum computers has been presented in \cite{Havl_ek_2019}.
Quantum simulators have become available to the scientific community only in recent years, and there is not a large literature on QML techniques applicable to these typologies of QC, especially in the case of neutral atoms. Since these type of platforms offers the possibility to create lattices of qubit defined by the user, they have the potential to efficiently implement kernels for graph classification. In this context, the QEK defined by Henry et al.~\cite{henry2021quantum} can extract features from graphs depending on their connectivity and structure. However, the authors validated the method using only classical emulation of quantum systems, leaving a full and optimized implementation on real quantum simulators for the future. A further generalization of this approach was developed in \cite{mernyei2022equivariant}, while other works explored the theoretical foundation of geometric QML \cite{skolik2023equivariant, Larocca_2022}. In \cite{Albrecht_2023}, the QEK has been validated using a 32 qubit simulator with limited connectivity. However, the used Quantum Processing Unit (QPU), besides having a low qubit number, is also restricted to triangular lattice, limiting the embedding flexibility.\\
In this context, classical ML methods have been developed to embed general graphs into the neutral atom simulator's register\cite{vercellino2022neural,10196840}. The goal of this work is to leverage the cited QEK and embedding method in order to implement an optimized procedure to classify for the first time a dataset composed of graphs with up to 256 nodes and arbitrary connectivity using the Aquila simulator.
%They have been used (with some proper modifications) for the preprocessing phase of the procedure we implemented, as described in detail in Section \ref{subsec:preprocessing}.

%% file: src/03_methodology.tex
\section{Methodology}\label{sec:methodology}
% While in the previous chapter, the proposed quantum evolution kernel is proposed and discussed, in this chapter its effectiveness is proven through experiments involving both emulation on a classical machine and simulation on a real quantum computer.
% The intrinsic difficulties of working with graph data are shown, in particular the preprocessing procedure involved in mapping to a quantum register.
% Later, the main discussion topic becomes hyperparameter tuning, in a double fashion: support vector machine parameter and quantum parameter, in the form of the waveform parameter and the necessary number of measurements.
% Several operative choices were taken during the work and are discussed in the following.
% In the end, the results are discussed, comparing the method with a classical kernel.
In this work, we implement and execute the QKE defined in ~\cite{henry2021quantum} on a real quantum simulator and use it to classify proteins from the \texttt{PROTEINS} dataset with a SVM. Before going into the details of the implementation, we will provide an overview of the overall procedure.\\
The QKE we are considering is composed of several layers of alternating parameterized constant pulses. In detail, for a given graph $\mathcal{G} = (\mathcal{V}, \mathcal{E})$ we define the mixing Hamiltonian $\hat{\mathcal{H}}_\mathcal{G}$ and the Ising Hamiltonian $\hat{\mathcal{H}}_\theta$ as,
$$\hat{\mathcal{H}}_\mathcal{G} = \sum_{(i,j) \in \mathcal{E}} \hat{\sigma}_i^z \hat{\sigma}_j^z$$
$$ \hat{\mathcal{H}}_\theta = \theta\sum_{i \in \mathcal{V}} \hat{\sigma}_i^y $$

Both Hamiltonians can be natively implemented in available neutral atom simulators by applying global Rabi pulses. Considering two layers, the parameter set 
$$\Lambda = \{\theta_0, t_0, \theta_1, t_1, \theta_2\}$$
fully defines the Hamiltonians and, therefore, the time evolution of the system.

\begin{figure}
    \centering
    \includegraphics[width=0.48\textwidth]{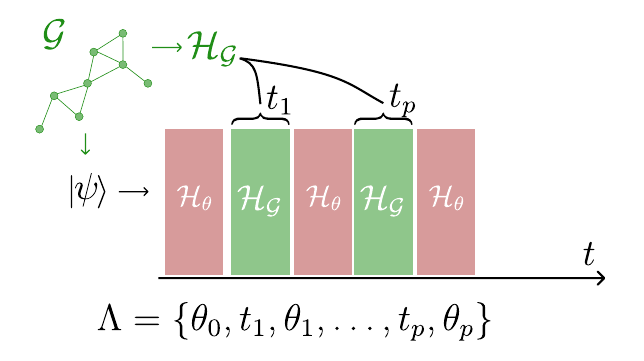}
    \caption{Layered time evolution, highlighting the order of application of the Hamiltonians and the parameters $\Lambda$.}
    \label{fig:layered approach}
\end{figure}

For a given $\Lambda$, the full \textit{hybrid quantum-classical}  procedure is detailed, summarized in Figure \ref{fig:layered approach}:
\begin{enumerate}
    \item \textbf{Quantum algorithm}: for every graph sample $\mathcal{G}_i$ and its position, a quantum register is instantiated and let to evolve following the Hamiltonian $\mathcal{H}_{\Lambda}$ defined by the parameters $\Lambda$.
    A set of measurements $M_{\mathcal{G}_i}$ is then obtained.
    \item \textbf{Classical post-processing}: a probability distribution is obtained from each $M_{\mathcal{G}_i}$ by computing the corresponding interaction energy.
    The output of this step is a set of probability distributions $\mathcal{P}_{\mathcal{G}_i}$ for each graph.
    \item \textbf{Machine Learning algorithm}: the set of $\mathcal{P}_{\mathcal{G}_i}$ is used to compute a kernel function between each graph employed in the training of the SVM. Finally, the optimal SVM hyperparameters have been selected through a K-fold cross-validation.
\end{enumerate}

After the Hamiltonian evolution, we sample the final state $\ket{\psi_f}$ measuring an observable $\hat{\mathcal{O}}$ with $\{\ket{o_i}, \dots \ket{o_K}\}$ eigenstates. From that we build a probability distribution

$$\mathcal{P}_{\mathcal{G}}^{\hat{\mathcal{O}}} (\Lambda) = (p_1, \dots p_K), \,\, \, p_i = \lvert \braket{o_i | \psi_f}\rvert^2$$

Given two probability distribution $\mathcal{P}$ and $\mathcal{P}'$, the Jensen-Shannon divergence is defined as:
$$JS (\mathcal{P}, \mathcal{P}') = H\left (\frac{\mathcal{P} + \mathcal{P}'}{2}\right) - \frac{H (\mathcal{P}) + H (\mathcal{P}')}{2}$$
where $H (\cdot)$ is the Shannon entropy of a probability distribution, defined as 
$$H (\mathcal{P}) = - \sum_k p_k \log{p_k}$$
The image of $JS (\cdot, \cdot)$ is the closed set $[ 0, \log{2}]$, and reach the maximum value ($\log{2}$) when the distribution have disjoint support. 
Finally, the graph kernel function of two graphs, $\mathcal{G}$, $\mathcal{G}'$ and their associated probability distributions $\mathcal{P}$ and $\mathcal{G}'$ is defined as:
\begin{equation}   
\mathcal{K}_{\mu}\left (\mathcal{G}, \mathcal{G}'\right) = e^{-\mu\, JS (\mathcal{P}, \mathcal{P}')}
\label{eq:kernel}
\end{equation}

with image in $[2^{-\mu}, 1]$. The $\mu$ hyperparameter can be optimized with the pulse parameters $\Lambda$, allowing the kernel to take value in a wider range.\\
As described in Section \ref{subsec:emu}, we used a Bayesian optimization procedure to obtain the best performing $\Lambda$. For this purpose, we use a subset of \texttt{PROTEINS} for which the algorithm can be classically emulated with reasonable time resources and time. 
Once defined the optimal $\Lambda$, we use quantum resources, namely the Aquila neutral atom simulator, for a larger subset of the considered dataset and compare the results of the classification.\\
In the following Sections, we give the details of each step of the procedure.

\section{Dataset and Preprocessing}

\texttt{PROTEINS} is one of the most used datasets for benchmarking graph machine learning algorithms.
Proteins are the perfect candidate to be modeled as graphs since they are macromolecules consisting of amino acids chain, disposed in a 3-dimensional space.
In fact, a graph $\mathcal{G} = (\mathcal{V}, E)$ is obtained by modeling each amino acid as a node $v \in \mathcal{V}$, and an edge between amino acids if they are less than $6 \,\AA$ apart in space.
Each protein comes with a binary label that describes whether it is an \textit{enzyme} or not.
One of the main problems when working with graph data is the absence of homogeneous data representation.
In~\cite{nr}, the authors tried to overcome this problem by collecting and uniforming graph data representation.
It comes in the form of different \texttt{.txt} files, each describing graphs, nodes, edges, and labels.
The dataset consists of 1113 graphs, divided into 663 enzymes and 450 non-enzymes.
Some overall statistics on the dataset are shown in Table \ref{tab:overall_graph_statistics}.
\begin{table}[ht]
    \centering
    \begin{tabular}{c|c}
        \hline
         Overall graph statistics \\
        \hline
        min number of nodes & $4$ \\
        avg number of nodes & $39.5$ \\
        max number of nodes & $620$ \\
        min number of edges & $5$ \\
        avg number of edges & $72.81$ \\
        max number of edges & $1049$ \\
        \hline
    \end{tabular}
    \caption{Overall statistics of the \texttt{PROTEINS} dataset.}
    \label{tab:overall_graph_statistics}
\end{table}

\begin{figure}[t]
    \centering
    %\hspace*{-1.5cm} 
    \includegraphics[width=1.\linewidth]{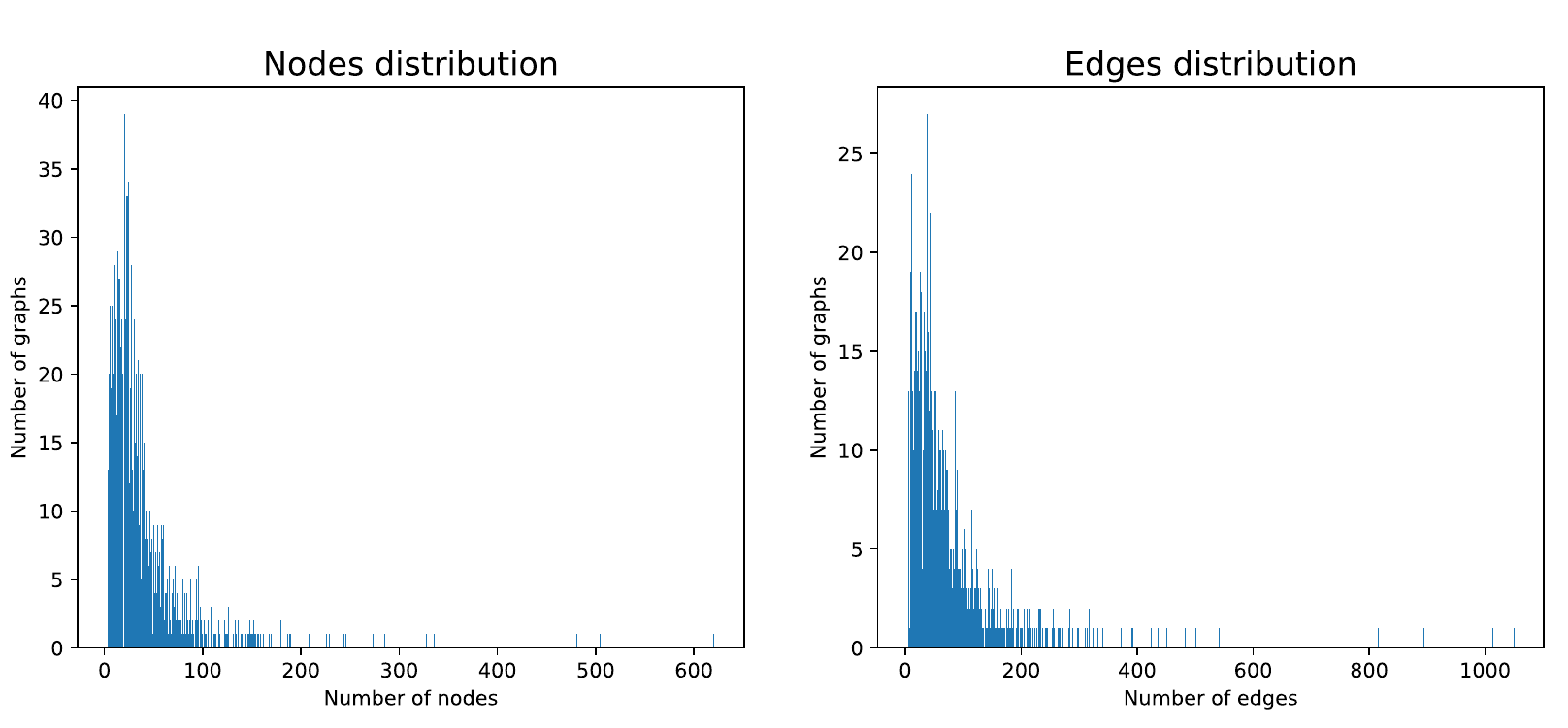}
    \caption{Nodes and edges distribution in \texttt{PROTEINS} dataset.}
    \label{fig:nodes_edges_distribution}
\end{figure}

In order to fit the graph on a quantum register, it should be limited to a maximum of 256 nodes (i.e.,
the maximum number of physical qubits available on Aquila).
By plotting the node number distribution, one can notice that most of the graphs contain less than 100 nodes.
Details on the distribution of nodes and edges are shown in Figure \ref{fig:nodes_edges_distribution}.

Since emulation time on classical machines grows exponentially with the number of qubits (i.e.,
the graph nodes) the dataset has been reduced, producing two different but overlapping datasets:
\begin{itemize}
    \item \texttt{PROTEINS256}: limited to graphs with a maximum of 256 nodes (Aquila current limit, corresponding to the number of physical qubits).
It consists of 276 graphs.
    \item \texttt{PROTEINS12}: limited to graphs with a maximum of 12 nodes.
The number of nodes is selected based on a qualitative benchmarking of emulation time.
It consists of 143 graphs.
\end{itemize}
It is important to note that both datasets are also limited by the constraint of being representable as UD graphs~\cite{clark1990unit}, as detailed later.
\subsection{Data preprocessing}\label{subsec:preprocessing}
As described in Section \ref{sec:methodology}, computing the probability distribution associated with each graph requires measurements from a quantum state that evolves following a time-dependent Hamiltonian (i.e.,
the alternation of mixing Hamiltonian $\mathcal{H}_\theta$ and Ising Hamiltonian $\mathcal{H}_\mathcal{G}$) that depends on the graph topology.
Given a graph, two methods can be implemented: 
\begin{enumerate}
    %\item Simulating the Hamiltonian without explicitly mapping to a register composed of atoms, in this way positions don't require to be approximated or computed. The Hamiltonian is built directly from the definition and emulated using classical machines. In the current settings, this method is not applicable for execution on real hardware.
    \item In~\cite{henry2021quantum} a simple rescaling method such that a minimum atom distance of $5 \, \mu m$ is applied. However, this approach strongly limits the number of graphs embeddable in a 2D register with the hardware constraints.

    \item Find the UD representation of the graph.
This way, the topology of the interaction of neutral atoms machine directly corresponds to the Ising Hamiltonian.
This method is the only one that is applicable to real quantum hardware without approximating  the Hamiltonian.
\end{enumerate}
Since the final objective of this work is to assess the quality of the methods on real hardware, the second method is the only one that allows the execution on a quantum computer.
As mentioned before, the first step is to embed the dataset's graphs into the register, which means finding the related UD graphs. For this, the DEN model has been used as a basis. All details about UD embedding with DEN can be found in \cite{vercellino2022neural,10196840}.\\
Some additional constraints had to be considered for the embedding into the Aquila register.
The neural-enhanced framework mentioned above allows further generalization, thanks to its flexibility.
In particular, these are the additional constraints of the chosen platform for the CUDG (Constrained Unit Disk Graph) problem:
\begin{itemize}
    \item \textit{Register area constraint}: the atoms' position should belong to a rectangle of size $75 \mu m \times 76 \mu m$.
    \item \textit{Row spacing constraint}: atoms should be positioned in discrete rows, with $4 \mu m$ spacing in between.
\end{itemize}
The last constraint is specific to Aquila, it has no correspondence with other neutral atom platforms, and is related to how the Aquila scheduler loads the atoms into the register.
The first constraint can be easily modeled by modifying the multiplier of the \textit{tanh} activation function, so that vertexes coordinates belong to the specific register's dimensions. The row spacing constraint is instead implemented through an approximation layer that adjusts the position in discrete rows, and by adding a new component to the Embedding Loss Function (ELF) that penalizes pairs that are vertically closer than 4 $\mu m$. The ELF, defined in \cite{10196840}, is designed to have a global minimum value for each feasible embedding found by the DEN model.\\
Finally, the UD constraints are then re-verified, in order to produce a feasible embedding.
\subsection{Embedded dataset}
The embedded datasets are produced by first filtering on the number of nodes and then employing DEN as described above, which reduces the number of graphs by filtering out graphs that cannot have a feasible UD embedding.
The characteristics of the generated datasets are presented in Table \ref{tab:final_dataset_number}.

\begin{table}[h]
    \centering
    \begin{tabular}{c|ccc}
                     & Number of graphs & \#Class 1 & \#class 2 \\
         \hline
         PROTEINS12  & 143 & 33 & 110 \\
         PROTEINS256 & 276 & 108 & 168 \\ 
         \hline
    \end{tabular}
    \caption{Final dataset composition.}
    \label{tab:final_dataset_number}
\end{table}

\begin{figure}[h]
    \centering
    \begin{subfigure}[b]{0.49\textwidth}
        \includegraphics[width=\textwidth]{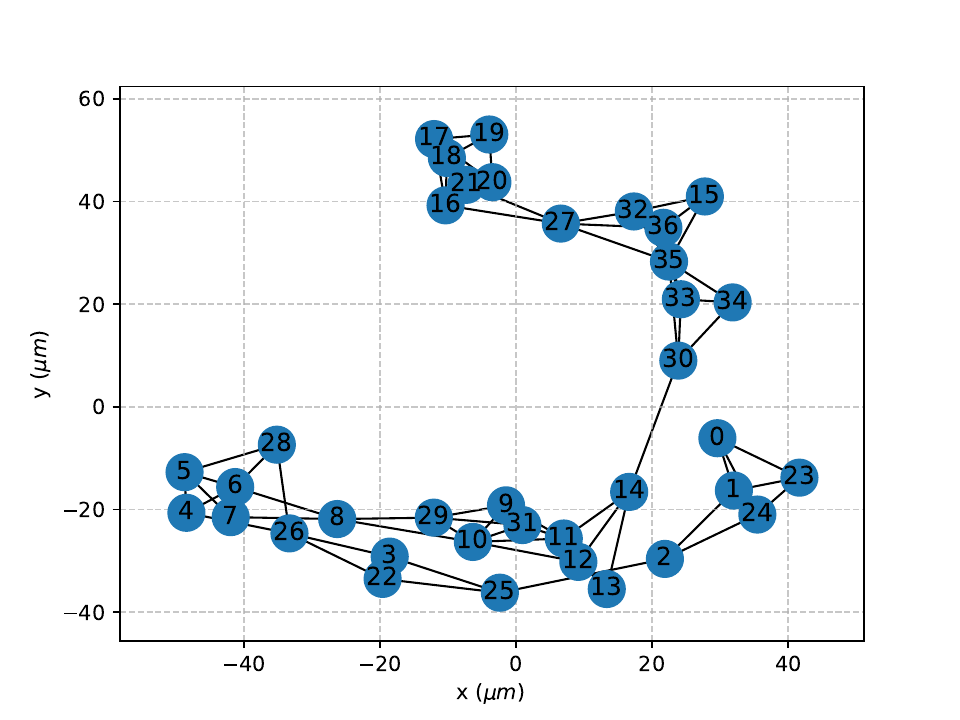}
    \end{subfigure}
    \begin{subfigure}[b]{0.49\textwidth}
        \includegraphics[width=\textwidth]{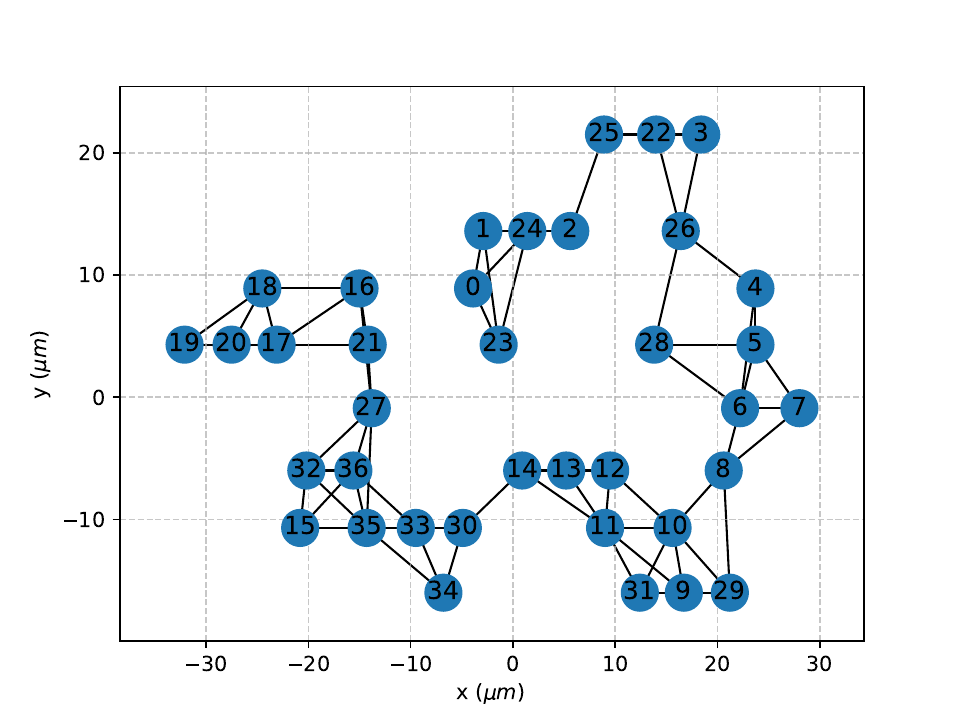}   
    \end{subfigure}
    \caption{example of initial not-UD graph (top) and the associated UD embedding produced by the modified DEN model (bottom). The discrete row constraint of Aquila is visible.}
    \label{fig:graph_ud_embedding}
\end{figure}

An example of a starting graph and the generated feasible embedding is presented in Figure \ref{fig:graph_ud_embedding}.

Unfortunately, the DEN model does not guarantee convergence, since it is an approximation method employing neural networks.
Not all graphs admit a UD representation, and this problem is more evident when the position space is limited to $\mathbb{R}^2$.
However, the number of embedded graphs is sufficient to make a preliminary analysis of the QEK benchmark.
It is important to note that the reduced dataset (i.e.,
\texttt{PROTEINS12}) presents a high-class imbalance.
Even if \texttt{PROTEINS256} does not present the issue to the same extent (class 1, representing $39 \%$ of the total dataset, while in \texttt{PROTEINS12} only the $23 \%$),  in the following sections some methods to overcome the imbalance are analyzed and then applied.

\section{Emulation on classical Hardware}\label{subsec:emu}
Classical emulation of the quantum evolution is performed using \textit{Bloqade}~\cite{bloqade}, a framework written in \textit{Julia}~\cite{julia} developed by QuEra for experimenting and interacting with their neutral atom quantum simulator.
Through its ecosystem, it supports emulating quantum systems, measuring different observables (even user-defined), and interacting with neutral atom hardware by validating and creating a representation for the Hamiltonian that can be sent to Aquila and run.
In particular, the used libraries are:
\begin{itemize}
    \item \textit{BloqadeODE}: which contains all the code for defining and emulating the time evolution of a quantum system by solving the Schrödinger equation.
    \item \textit{BloqadeSchema}: which contains all the functions for validating the Hamiltonian, in particular for checking the feasibility of the atoms' position, and contains the function for smoothing the waveform in the case of piecewise protocols, according to the bandwidth of the lasers.
\end{itemize}

\begin{figure}[t]
    \centering
    \includegraphics[width=0.49\textwidth]{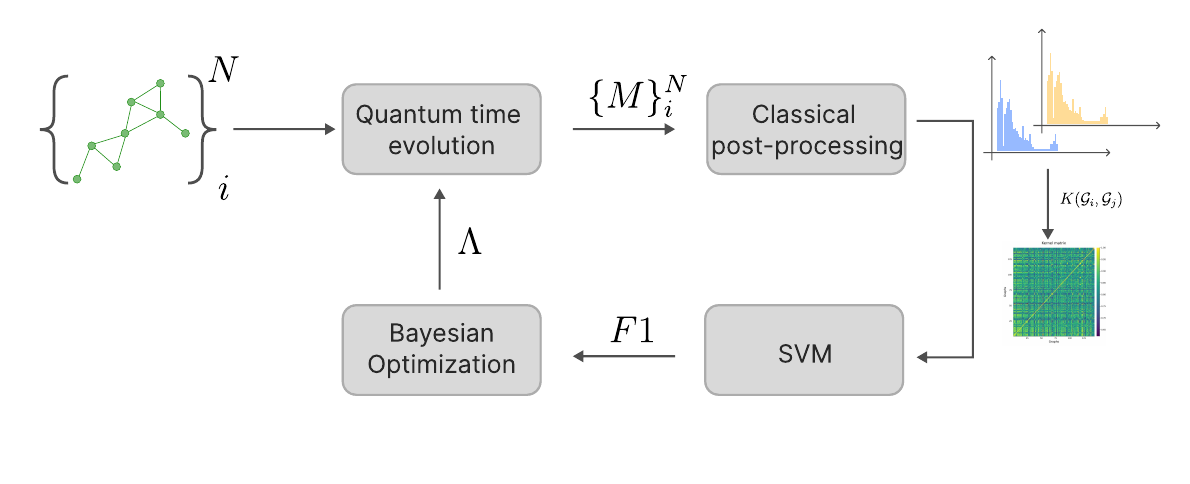}
    \caption{The full Hybrid quantum-classical emulation approach.
}
    \label{fig:hybridquantumclassicalapproach}
\end{figure}

As described before, the emulation step is very computationally onerous, but still necessary, for two main reasons:
\begin{enumerate}
    \item Confirm the validity of the method before executing on quantum resources.
    \item Each iteration of the Bayesian optimization algorithm, as described later, requires a total number of measurements (also called shots) $N_{shots} = n_{shots} \times n_{graphs} = 276 \cdot 10^3$ (where $n_{graphs} = 276$ is the number of graphs and $n_{shots} = 10^3$ is the number of shots per graph) to run on the quantum computer, which is too resource-expensive.
This step is required to find the waveform parameters $\Lambda$.
\end{enumerate}
The full \textit{hybrid quantum-classical} emulation approach is summarized in Figure \ref{fig:hybridquantumclassicalapproach}.

After the steps reported in Section \ref{sec:methodology}, the \textbf{Bayesian Optimization} phase receives the new pair $ (\Lambda, f (\Lambda)$ and updates the posterior probability distributions.
The acquisition function produces a new set of parameters $\Lambda$ for the next iteration.\\
In the following, the experimental settings and results are discussed.
\subsection{Probability distribution}
As described in Section \ref{sec:methodology}, the core of the method is the computation of a graph kernel as the distance between two probability distributions representing the graphs.
%In the following, the operative approach to computing the energy distribution is discussed.
Aquila allows measurements only in the computational basis.
The \textit{energy observable} is then computed by post-processing the set of measurements.

Given a graph $\mathcal{G}$, and a set of measurements $M$:
\begin{gather*}
    M = \{ m_1, \dots, m_k \}, 
    m_i = [n_1, \cdots ,n_n], \\ n_i = \left \{\begin{array}{cc}
     1 & \text{if atom i is measured in } \ket{g}  \\
     0 & \text{if atom i is measured in } \ket{r}
\end{array}\right .
\end{gather*}
The used number of shots is $k \leq n_{shots}$ because the measurement process can fail (due to an erroneous register initialization), in which case the shot is discarded.
Each measurement $m_i$ is composed by a \textit{bitstring} of length $|\mathcal{V}|$ (i.e.
the number of nodes/atoms).
%It is important to notice %that the single bit is %flipped with respect to %how is expected.
%The value 1 is measured %when the atom is %presented into the ground %state.
%Since is common to find %in quantum algorithms the %measured value is equal %to the logical state, a %bit-flipping operation is %applied to each %measurement.
For each measurement $m_i$ we can associate an energy:
$$e_i = \sum_{j < k} V_{jk}n_jn_k$$
By repeating the process for every correctly initialized shot, a set of energies related to each single graph is computed, indicated as $E_{\mathcal{G}_i}$.
In order to have comparable distributions, they should have the same support.
For this reason, a binning procedure is used.

\begin{figure}[t]
    \centering
    %\begin{subfigure}[b]{.4\textwidth}
    %    \includegraphics[width=\textwidth]%{images/ch4/97.pdf}
    %    \caption{Energy distribution and the %associated graph}
    %\end{subfigure}
    \begin{subfigure}[b]{.4\textwidth}
        \includegraphics[width=\textwidth]{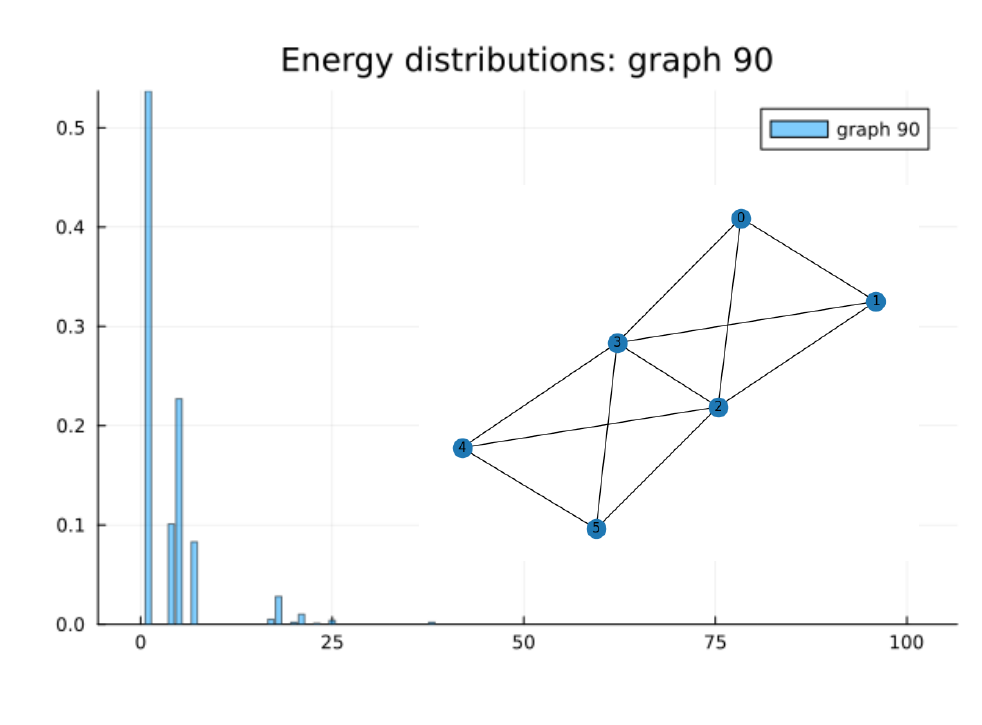}
        \caption{Energy distribution and the associated graph.}
    \end{subfigure}
    \begin{subfigure}[b]{.4\textwidth}
        \includegraphics[width=\textwidth]{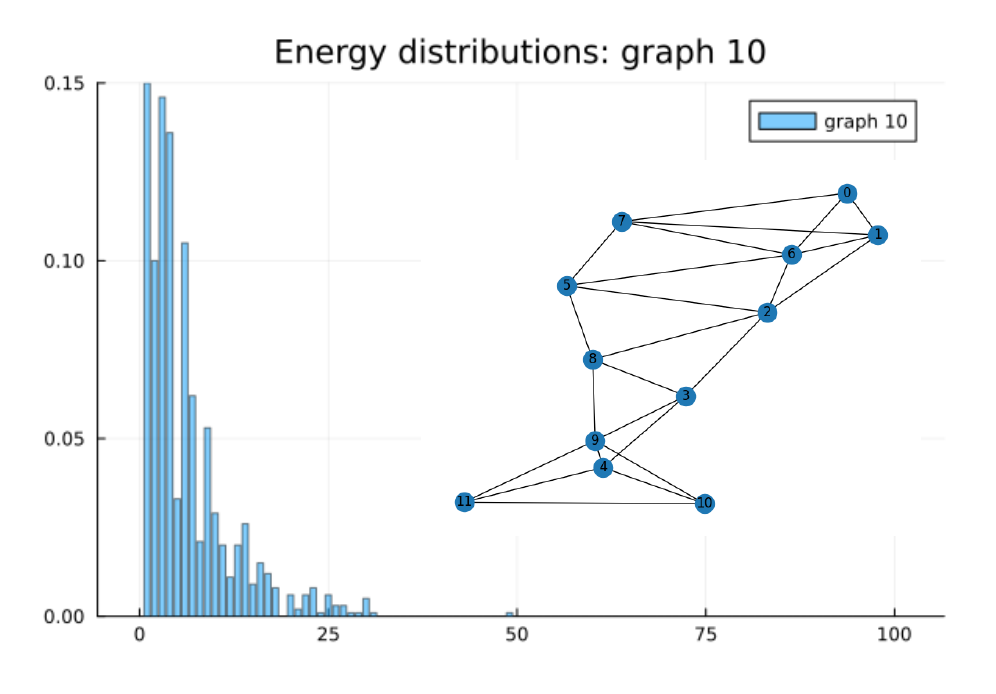}
        \caption{Energy distribution and the associated graph.}
    \end{subfigure}
    \begin{subfigure}[b]{.4\textwidth}
        \includegraphics[width=\textwidth]{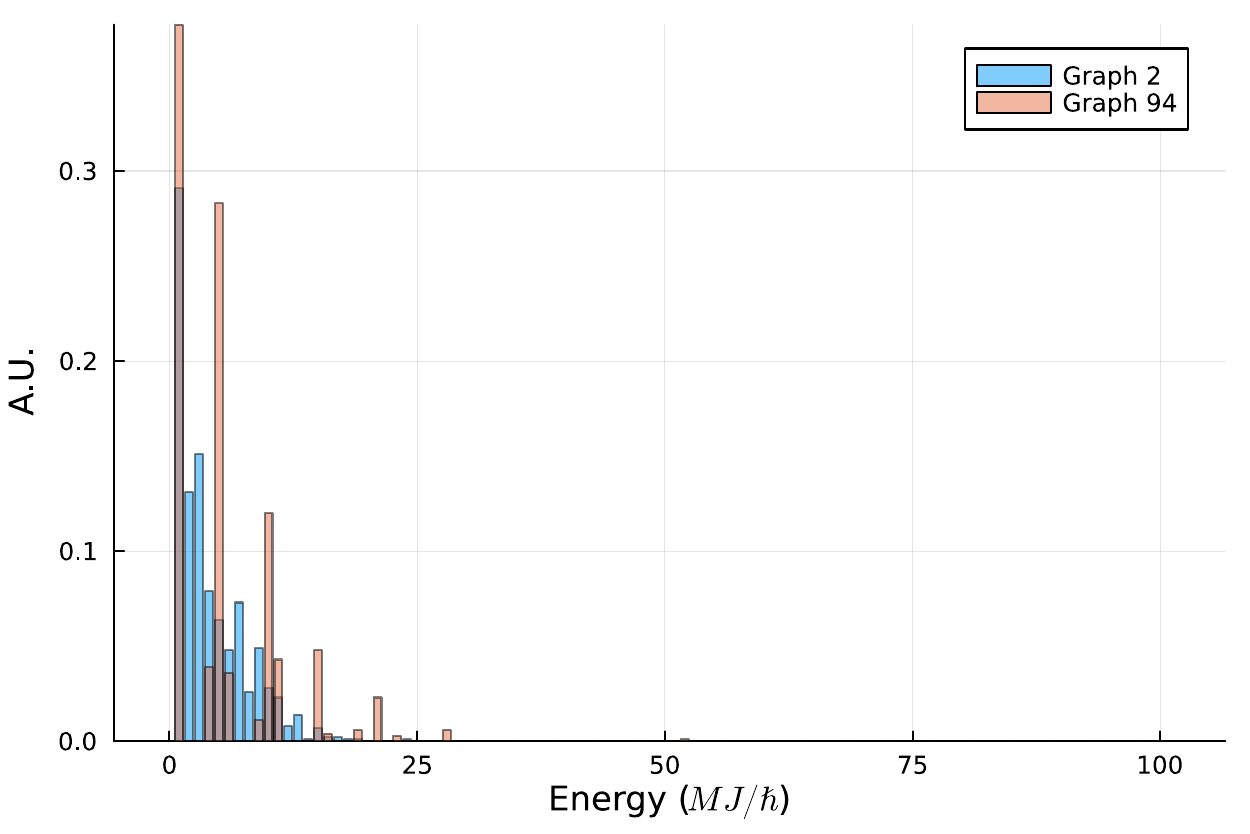}
        \caption{Comparison of the energy distributions of two graphs belonging to different classes.}
    \end{subfigure}    
    \caption{Example of obtained energy distributions for two different graphs.}
    \label{fig:energy_distribution}
\end{figure}
Once every energy distribution is computed, these two values are calculated:
$$ e_1 = \min_i \{ \min_{e_j} E_{\mathcal{G}_i}\} \,\,,~~e_2 = \max_i \{ \max_{e_j} E_{\mathcal{G}_i}\}$$
% Where $e_1$ is the minimum of all measured energies, and $e_2$ is the maximum measured energy.
Where $e_1$ and $e_2$ are respectively the minimum and maximum measured energy.
Selecting 100 equal-sized bins in the interval $[e_1, e_2]$ allows to compute a probability distribution $\mathcal{P} (\mathcal{G})$.
Bins are normalized to sum up to 1 by dividing by the number of measurements $k$.
Examples of the probability distributions and the associated graphs can be found in Figure \ref{fig:energy_distribution}.

\subsection{Kernel estimation and training protocol}
\begin{figure}[t]
    \centering    \includegraphics[width=0.47\textwidth]{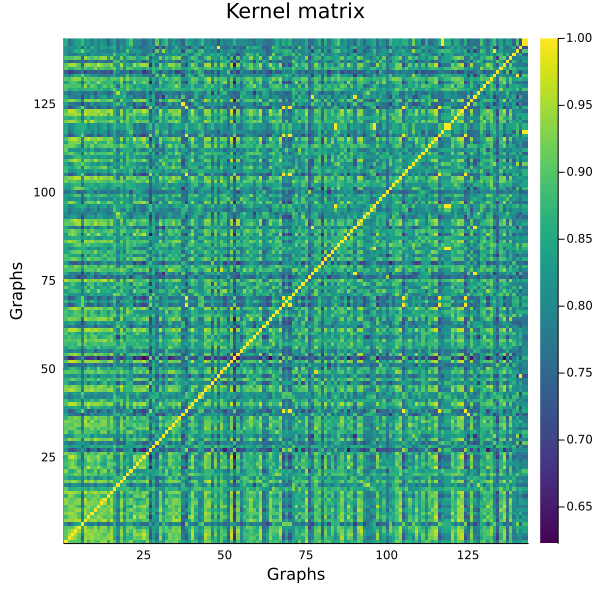}
    \caption{Kernel matrix of \texttt{PROTEINS12} dataset, with $\mu = 1$.}
    \label{fig:kernelPROTEINS12}
\end{figure}

Once the probability distribution is estimated, the kernel is computed as described in Eq.
\ref{eq:kernel}.
The only hyperparameter involved in this task is $\mu$.
Following~\cite{henry2021quantum}, $\mu$ is set equal to $1$.
Further hyperparameter tuning is possible, knowing that $\mu$ influences the image of the kernel.
With $\mu = 1$  $K (\cdot) \in \left[ \frac{1}{2}, 1 \right]$, while for bigger values the image increases, and in the limit of $\mu \to \infty$ reaches $\left[ 0, 1 \right]$.
Once the QEK is estimated, it is organized in a kernel matrix $K$.
The entries are organized as:
$$K_{i, j} = K (\mathcal{G}_i, \mathcal{G}_j)$$
The kernel matrix expresses the similarity between pairs of graphs and can be visualized graphically using a heatmap, as in Figure \ref{fig:kernelPROTEINS12}.
As we can notice, there is a brighter zone in the bottom-left corner associated to higher kernel values.
This group is composed of graphs labeled as \textit{enzymes}, demonstrating qualitatively that the quantum kernel is able to find structures in data.
The kernel matrix is then used to train a SVM.
To evaluate the approach and tune the SVM hyperparameters, validation data are generated using a K-fold cross-validation scheme (with $K=10$), in combination with a \textit{grid search} to find the optimal parameters.
The hyperparameters tuned in this phase are:
\begin{itemize}
    \item $C$: 100 points $[10^{-4}, 10^4]$, logarithmic scale.
    \item $w$: 30 pairs $[1, x_i]$ with $x_i$ in $[1, 1000]$.
\end{itemize}

For every possible combination of the parameter, 10 models are trained on 9 fold of the data, and validated on the remaining.
Each model score is then collected and averaged, producing single metrics used for evaluation.

The $C$ hyperparameter expresses the possibility of the classifiers of making misclassification errors during the training procedure.
This improves the generality of the solution and prevents overfitting.
The additional pair of class weights $w$ is used to prevent the model from predicting always the majority class.
They act as a penalty for misclassifying the minority class.
\subsection{Bayesian optimization of Waveform parameters}
%The architecture follows the one proposed in~\cite{henry2021quantum} and detailed in Section \ref{sec:methodology}.
In terms of pulse sequence, the selected approach consists of three pulses of a single qubit drive (i.e.,
the mixing Hamiltonian) alternated by two free evolutions (no pulse).
The amplitude of the Rabi drive is kept constant and equal to the maximum amount reachable by the QPU ($\Omega_0 = 15.8 \, rad/\mu s$, in order to reduce the overall evolution time, and consequently to reduce the noise sources).
As discussed in Section \ref{sec:methodology}, the full dynamics can be realized using a time-dependent Hamiltonian with a $\Omega (t)$ governed by a pulse that alternates values where the drive is on (corresponding to the mixing Hamiltonian) with a period where the drive is off, corresponding to free evolution of the quantum state.
Given this setup, we can redefine the parameter set as:
$$\Lambda_{BO} = \{\tau_0, t_0, \tau_1, t_1, \tau_2\}$$
Where $\tau_i$ represents the Rabi drive duration (and consequently the angle of rotation), while $t_i$ represents the free evolutions.
It is possible to retrieve the angle of rotation of the mixing Hamiltonian by remembering that $\theta_i = \Omega_0 \, \tau_i$.\\
Given a set of parameters $\Lambda$,
the training protocol described before returns the SVM hyperparameter set that reaches the best average score among the 10 folds.
However, the pulse parameter has to be tuned, in order to find the pulse that extracts a good amount of knowledge (i.e.,
has the maximum F1 score).
Formally this can be seen as a maximization problem, finding the $\Lambda$ that maximizes $f: \mathcal{X} \to \mathbb{R}$, where $\mathcal{X}$ is the space of the parameters $\Lambda$, and $f$ is the average selected score (F-1 in this case) among the 10 folds of the \texttt{PROTEINS12} dataset.
However, the evaluation of $f$ is costly, since it consists of training a model and computing  the state several times for each graph, and common heuristic methods are not feasible.
Bayesian optimization~\cite{frazier2018tutorial} is particularly well suited for optimizing \textit{black-box functions}, in which evaluating different sets of parameters can be costly.
The framework is made of two components:
\begin{itemize}
    \item \textit{Surrogate function} $\Tilde{f}$: approximates the costly objective function starting from a set of evaluations of $f$.
In Bayesian optimization, it is usually called \textit{prior} and reflects the knowledge we have on the function $f$.
    \item \textit{Acquisition function} $\alpha (x)$: indicates where to sample the next point to evaluate $f$.
\end{itemize}
It is common to select the surrogate function as a Gaussian Process $\mathcal{GP}$, characterized by a mean value $\mu$ and a covariance matrix $\Sigma$. The Gaussian Process so defined has a probability density function of the form:
$$
P (X) = \frac{1}{\sqrt{ (2\pi)^k |\Sigma|}}\cdot \exp{ \left (-\frac{1}{2} (X - \mu)^T \Sigma^{-1} (X-\mu)\right)}
$$
where $k$ is the variable's dimensionality. The selected mean $\mu$ is Normal distributed as $\mathcal{N} (0, 10)$, while the covariance function is selected as the \textit{Isotropic $5/2$ Matèrn Kernel}, with length scale $\ell = 10$ and signal standard deviation $\sigma = 10$:

\begin{align}
K (x, x') = & \sigma^2 (1 + \sqrt{5 |x - x'|/\ell + 5|x - x'|^2/ (3\ell^2)}) \nonumber\\
\ & exp{\left (-\sqrt{5|x - x'|/\ell}\right)}
\end{align}

\begin{figure}[t]
    \centering
    \begin{subfigure}[b]{.48\textwidth}
        \includegraphics[width=\textwidth]{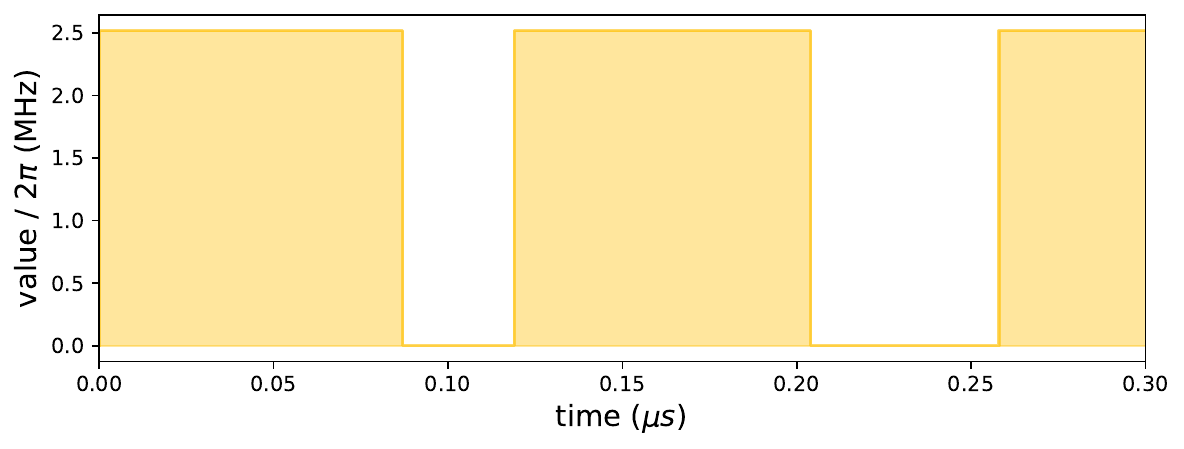}
        \caption{$\Lambda_a$ parameter set.}
    \end{subfigure}
    \begin{subfigure}[b]{.48\textwidth}
        \includegraphics[width=\textwidth]{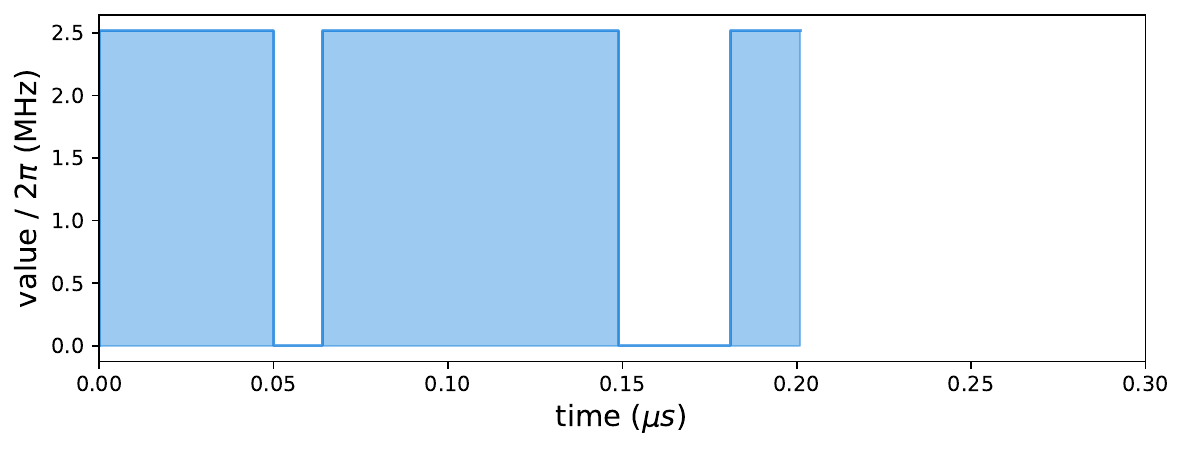}
        \caption{$\Lambda_{BO}$ parameter set.}
   \end{subfigure}
    \caption{Waveforms  comparison.}
    \label{fig:waveform_parameters}
\end{figure}

The acquisition function is the commonly employed \textit{Upper Confidence Bound}, $\alpha (x) = \mu (x) + k\sigma (x)$, where $k$ is a hyperparameter that expresses the trade-off between exploration ($\mu$ parameter) and exploitation ($\sigma$ parameter).
The kernel hyperparameters (length scale and signal standard deviation) are optimized every 10 iterations, using Maximum A Posteriori (MAP) estimate.
The maximum number of iterations is set to 50.
In addition, some constraints are applied to the optimized variables:
$$
\begin{array}{cc}
     \tau_0 + t_0 + \tau_1 + t_1 + \tau_2 & < 500 ns \\
     t_i > 5 ns & \forall i \in [0, 1] \\
     \tau_i > 5 ns & \forall i \in [0, 1, 2] \\
\end{array}
$$
The first constraint limits the total execution time, preventing decoherence and consequently reducing both the simulation and classical emulation time, while the second and third constraints are related to the finite bandwidth of the optical components.
%The first constraint is related to the \textit{decoherence time} of the machine and hence pushing for a reduction in the emulation time, while the second and third are related to the time resolution of the laser fields.
The waveform obtained by the entire process on the \texttt{PROTEINS12} is then compared with the one obtained in~\cite{henry2021quantum}, named here for simplicity $\Lambda_a$, in Figure \ref{fig:waveform_parameters}.
%\begin{figure}[h]
%    \centering
%    \includegraphics[width=\linewidth]{images/ch4/waveform_comparison.pdf}
%    \caption{Parameters}
%    \label{fig:waveform_parameters}
%\end{figure}

\begin{figure}[t]
    \centering
    \includegraphics[width=.49\textwidth]{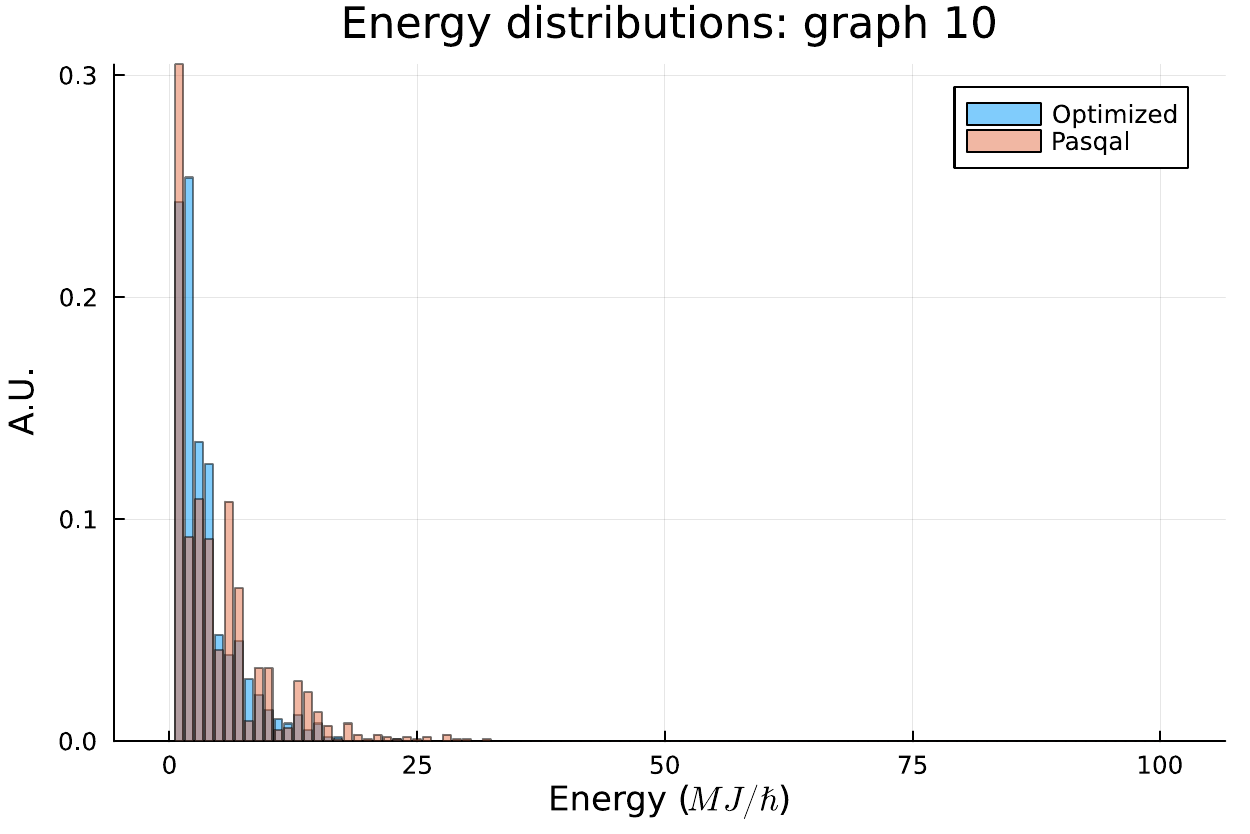}
    \caption{Comparison of the energy distributions obtained using two different sets of waveform parameters on the same graph.}
    \label{fig:distributionParameter}
\end{figure}

In general, the two waveforms are similar.
Both are characterized by short free-evolution and longer periods of  $\mathcal{H}_{\theta_i}$.
However, the total duration of $\Lambda_a$ waveform is $317 \, ns$, in contrast to the $201 \, ns$ of the one found in this work.
This difference of $\simeq 100 \, ns$ has one major benefit: a shorter protocol is expected to produce less noisy results, allowing quantum simulations to be more similar to classical emulation and to produce higher quality results, as showed in the following.\\
Another relevant difference consists in the mixing Hamiltonian duration: for $\Lambda_a$, the three mixing Hamiltonian have similar duration ($87 \, ns$, $84 \, ns$ and $72 \, ns$), while $\Lambda_{BO}$ is characterized by a long central pulse (comparable to the one in $\Lambda_a$ waveform, $85 \, ns$ corresponding to a rotation $\simeq 80^\circ$) and two shorter pulses of $50 \, ns$ and $20\, ns$, corresponding respectively to a rotation of $45^\circ$ and $18^\circ$.
It is possible to qualitatively compare the energy distribution of the same graph obtained with the two different $\Lambda$, as shown in Figure \ref{fig:distributionParameter}.

\subsection{Results on PROTEINS12}
The results obtained in the emulation experiment can be presented, in the form of \textit{F1 score}, \textit{Accuracy}, \textit{Precision} and \textit{Recall} metrics.
As a benchmark, the method is compared with the version using the $\Lambda_a$ parameter set, and with a classical algorithm, the \textit{Shortest Path graph Kernel} (SPK)~\cite{borgwardt2005shortest}.
The emulation platform is based on a \textit{Dell-XPS 15 7590}.
The results are presented in Table \ref{tab:proteins12-results}.
\begin{table}[htbp]
    \centering
    \setlength\tabcolsep{1pt}
    \begin{tabular*}{\columnwidth}{@{\extracolsep{\fill}} c|cccc}
         & F-1 (\%) & Accuracy (\%) & Precision (\%) & Recall (\%) \\
         \hline
         $\Lambda_a$ & 41.9 & 44.8 & 29.3 & 89.1 \\ 
         $\Lambda_{BO}$ & \textbf{46.3} & 52.4 & 37.1 & 85 \\ 
         Shortest Path kernel & 44.1 & \textbf{61.4} & 35 & 68 \\ 
         \hline
    \end{tabular*}
    \caption{Performance comparison of QEK using $\Lambda_a$ and $\Lambda_{BO}$ parameter sets with the SPK on \texttt{PROTEINS12}.}
    \label{tab:proteins12-results}
\end{table}
% commento sui risultati
\\
As expected, the performance scores of all the kernels on the reduced dataset are generally low due to the small size of the training sample, even if using a K-fold-cross validation approach with an high K ($K=10$).
However, the kernel computed using $\Lambda_{BO}$ outperforms the one obtained with $\Lambda_a$ by $\simeq 5 \%$ on the F1 score.
The improvement on the F1 score is reflected also in an higher accuracy (44$\%$ against $52.4 \%$) and precision ($29.3 \%$ against $37 \%$).
Notably, the implemented QEK method performs even better than the classical SPK in terms of F1 score, but with a smaller margin ($44.1 \%$ of the SPK vs.
the $46.1 \%$ of QEK). The same is true for precision and recall.
However, the accuracy of the classical model is higher, because of the high-class imbalance.
This shows how in this case the accuracy score can be misleading.

%% file: src/04_results.tex
\section{Simulation on Aquila: Results and Discussion} \label{sec:aquila}
We proceed to execute the proposed algorithm on the real noisy quantum simulator Aquila for both datasets.\\
The interaction with the hardware is ensure by the \textit{BloqadeHardware} library, which in particular provides support to properly formatting the defined Hamiltonian and to check the compliance with the hardware constraints.
Before delving into the results, it is important to report the number of shots that have been performed on the quantum hardware.
Each graph in the dataset is associated with a single Hamiltonian evolution, i.e., a single task for Aquila.
As per the classical emulation case (see Section \ref{subsec:emu}), the total number of measurement for the whole dataset is $n_{shots} \times N_{graphs} = 276 \cdot 10^3$ for each set of parameters $\Lambda$. In the following, we report the detailed results.

\subsection{Comparison of energy distributions}
\begin{figure}[t]
    %\centering
    %\begin{subfigure}[b]{.4\textwidth}
    %    \includegraphics[width=\textwidth]%{images/ch4/energy_distribution_14_simulated_vs%_emulated.png} 
    %    \caption{}
    %\end{subfigure}
    %\begin{subfigure}[b]{.4\textwidth}
    %    \includegraphics[width=\textwidth]%{images/ch4/energy_distribution_58_simulated_vs%_emulated.png}
    %    \caption{}
    %\end{subfigure}
    %\begin{subfigure}[b]{.4\textwidth}
        \includegraphics[width=.49\textwidth]{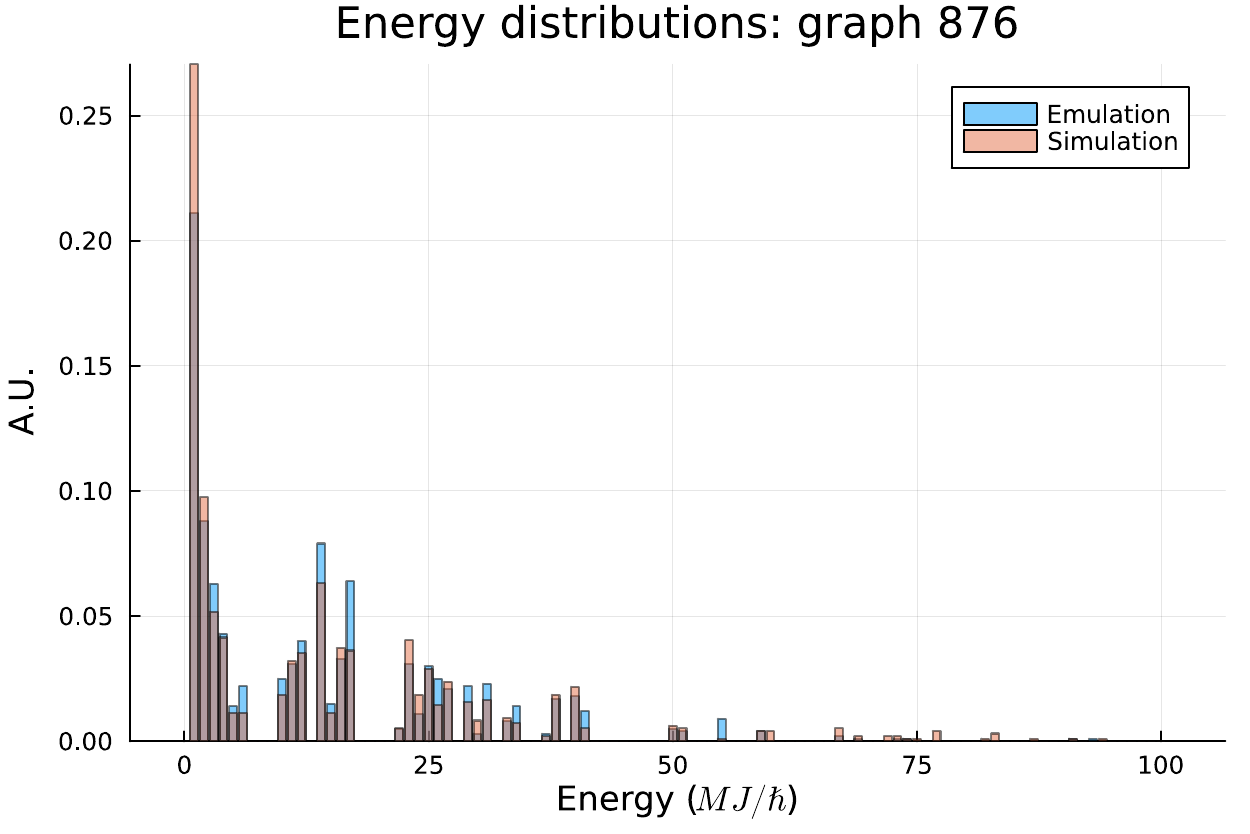}
        %\caption{}
    %\end{subfigure}
    %\begin{subfigure}[b]%{.4\textwidth}
    %    %\includegraphics[width=\t%extwidth]%{images/ch4/energy_distri%bution_63_simulated_vs_em%ulated.png}    
    %    \caption{}
    %\end{subfigure}
    \caption{Comparison between the emulated and simulated energy distributions a randomly selected graph.}
    \label{fig:ed_comparison}
\end{figure}

Before executing the QKE on the full dataset, a random subset is extracted and the results of the quantum simulation are compared to the one obtained with the classical emulation.
The comparison is done in terms of the obtained energy distribution since computing the statevector would require an exponential number of measurements.
 Figure \ref{fig:ed_comparison} shows the energy distribution comparison for an example graph.
The two distributions are similar, demonstrating a low impact of noise on the approach.
Notably, in simulation there is a higher probability to have very low energy configuration (i.e., almost zero atoms in the Rydberg state) than emulation. This is probably due to the coupling to the ground state, which is missing in the considered noiseless classical emulation.
At the same time, some higher energy configurations are present in the simulation.
These can be partially explained by the detection errors that affect atoms in the ground state $\ket{g}$ to be detected as in the Rydberg state $\ket{r}$, increasing the probability of detecting a double-excited Rydberg state.
\subsection{Results on PROTEINS12}
The experiment is repeated with the same experimental setup presented before for the full \texttt{PROTEINS12}, in order to compare the classification of emulated and simulated QEK.
The experiment is repeated with both $\Lambda_{BO}$ and $\Lambda_a$ sets.
%The choice of simulating both parameters comes from the consideration that Henry et al.~\cite{henry2021quantum} did not simulate on a real machine, and they highlighted that an interesting way to continue their work would be to simulate on an actual neutral atoms machine.
%The second consideration is to highlight the differences in terms of the result for 2 protocols with different durations.
As reported in the Section \ref{subsec:emu}, the $\Lambda_{BO}$ protocol has a $33 \%$ shorter duration.
This difference in circuit duration can help to evaluate the effect of noise for longer protocols.
The result presented in Table \ref{tab:proteins12-results-simulation} confirms that the method is effective, even in the case of the real noisy simulation.

\begin{table}[h]
    \centering
    \setlength\tabcolsep{1pt}
    \begin{tabular*}{\columnwidth}{@{\extracolsep{\fill}} c|cccc}
         & F-1 (\%) & Accuracy (\%) & Precision (\%) & Recall (\%) \\
         \hline
         $\Lambda_a$ (Emulation) & 41.9 & 44.8 & 29.3 & 89.1 \\ 
         $\Lambda_{BO}$ (Emulation) & \textbf{46.3} & 52.4 & 37.1 & 85 \\ 
         SPK & 44.1 & \textbf{61.4} & 35 & 68 \\ 
         $\Lambda_a$ (Aquila)& \textbf{48} & 62.2 & 35.9 & 80.8 \\
         $\Lambda_{BO}$ (Aquila) & \textbf{49} & 63.7 & 39.5 & 78\\
         \hline
    \end{tabular*}
    \caption{Performance comparison of QEK using $\Lambda_a$ and $\Lambda_{BO}$ sets, in  emulated and simulated execution, with the SPK.}
    \label{tab:proteins12-results-simulation}
\end{table}

The simulation on the Aquila quantum computer with the optimized waveform parameters has the best result, outperforming both the chosen classical method and the emulated one (of about $3\%$ in the F1 metric).
This result is also reflected in the accuracy metric of $63.7 \%$, the highest among the considered methods.
Interestingly, also in the case of the $\Lambda_a$ parameter set, the simulation scores are higher than the emulated ones.
%In terms of noise, the longer protocol does not seem to interfere with the model performance, registering one of the best results among all the methods.
\subsection{Results on PROTEINS256}
Finally, the experiment using the full dataset has been conducted.
The results can be found in Table \ref{tab:proteins256-results}.

\begin{table}[htbp]
    \centering
    \setlength\tabcolsep{1pt}
    \begin{tabular*}{\columnwidth}{@{\extracolsep{\fill}} c|cccc}
         & F-1 (\%) & Accuracy (\%) & Precision (\%) & Recall (\%) \\
         \hline
         Aquila simulation & \textbf{65,6} & \textbf{65,4} & \textbf{55,1} & 86,6 \\ 
        SPK & 65,3 & 64,9 & 53,5 & \textbf{87,1} \\ 

    \end{tabular*}
    \caption{QEK and SPK performances on \texttt{PROTEINS256}.}
    \label{tab:proteins256-results}
\end{table}

% The model performances are good enough, with an accuracy of about $65 \%$, slightly better than the classical kernel.
In general, performance scores are higher (about 20\% on the F1 metric) than the \texttt{PROTEINS12} case due to the larger size of the dataset and the higher class balance.
Even if a weighting training objective and a cross-validation approach with a high value of K are used, the number of training samples is not enough to provide reliable classification results.
Even in this experimental setting, the simulation of the QEK on Aquila with the optimized parameter has performances that are comparable to that of the classical SPK kernel. It is important to note that, has mentioned before, the $\Lambda_{BO}$ has been optimized only on the \texttt{PROTEINS12} dataset in emulation, due to limited computational and quantum resources available.
This could have had a detrimental effect on the results obtained.
Moreover, the larger difference between the QEK and the SPK for the \texttt{PROTEINS12} case with respect to the \texttt{PROTEINS256} one seems to experimentally corroborate the claim that quantum kernels are able to extract features more efficiently than classical ones using fewer training data \cite{Caro_2022}.

In order to save quantum resources and time in future works (Aquila is able to sample at an approximate rate of 10 Hz\cite{wurtz2023aquila} with an associated cost per shot), a statistical analysis can be performed on the number of shots required to accurately compute the energy distribution.
%The number of shots is directly related to the cost of the experiment, both in monetary cost and in time.
%This means that while an experiment with a single graph could last 10 seconds with 100 shots, with 1000 shots it would last 10 times more.
To this end, a sampling approach has been used.
From the whole set of measurements, a number $k$ of shots is drawn randomly and without replacement, simulating an experiment with $k$ number of shots.
Then, the same post-processing and cross-validation approach was used.
The results are shown in Table \ref{tab:proteins256-n_shots_analysis}.

\begin{table}[h]
    \centering
    \setlength\tabcolsep{1pt}
    \begin{tabular*}{\columnwidth}{@{\extracolsep{\fill}} c|cccc}
         number of shots & F-1 (\%) & Accuracy (\%) & Precision (\%) & Recall (\%) \\
         \hline
         10   & 59,3 & 63,1 & 52,9 & 71 \\ 
         100  & \textbf{65,1} & 64,4 & 53,8 & 87 \\ 
         1000 & \textbf{65,6} & 65,4 & 55,1 & 86,6\\ 
         \hline
    \end{tabular*}
    \caption{\texttt{PROTEINS256} performances against the number of shots.}
    \label{tab:proteins256-n_shots_analysis}
\end{table}

The overall performance does not degrade much when the number of shots is reduced to 100 (about 1\% average degradation), while it rapidly decreases when the number of samples is reduced to 10.

%% file: src/05_conclusion.tex
\section{Conclusion}\label{sec:conclusion}

This work demonstrated the ability to successfully implement a QML algorithm for graph classification on the Aquila simulator.
By mapping graphs from the \texttt{PROTEINS} dataset onto the register of Rydberg atoms using state-of-the-art ML methods and computing the Quantum Evolution Kernel, competitive classification performance was achieved compared to classical graph kernel methods like the Shortest Path Kernel.
An optimization procedure based on Bayesian optimization was used to find a set of waveform parameters for the quantum evolution that outperformed previously reported parameters found in literature.
When simulated on the noisy Aquila hardware, the optimized QEK achieved an F1 score of $65.6\%$ on the full \texttt{PROTEINS256} dataset, comparable to $65.3\%$ for the SPK.

Despite the limited number of qubits and noise present on current neutral atom quantum hardware, these results show the feasibility of implementing useful QML workloads on these systems.
As neutral atom architectures scale up to more qubits and improve in terms of fidelity and coherence times, their inherent connectivity mappings may provide advantages for graph analytics tasks.
%Further work is still needed to assess quantum advantage over classical methods, incorporate error mitigation, and explore other applications of neutral atom quantum simulators.
In summary, this study serves as an important first proof-of-concept demonstration for executing QML workloads for graph problems with hundreds of nodes and arbitrary connectivity on neutral atom quantum computing platforms.
As this emerging quantum hardware develops further, exploring the interplay of algorithms and hardware will be crucial to unlocking its potential advantages.